\documentstyle[aaspptwo]{article}

\begin{document}

\title{THE POLARIZED SPECTRUM OF THE FE II-RICH BAL QSO, IRAS 07598+6508}

\centerline{astro-ph/9505148}

\centerline{Submitted 1995 April 16, to appear approx. July, ApJ Letters}

\author{Dean C. Hines\altaffilmark{1} and Beverley J. Wills\altaffilmark{2}}
\affil{McDonald Observatory and Astronomy Department, University of Texas,
    Austin, TX 78712}
\altaffiltext{1}{Current Address: Steward Observatory, \\
University of Arizona, Tucson AZ 85721}
\altaffiltext{2}{Guest Observer with the International Ultraviolet Explorer.}

\begin{abstract}

We present IUE spectrophotometry and optical spectropolarimetry of
the ultraluminous, extreme FeII-emitting QSO IRAS 07598+6508.  We
find broad absorption troughs from high- and low-ionization species,
showing that this object is a member of the class of rare
low-ionization BAL QSOs.  Compared with non-BAL QSOs, the spectral
energy distribution is reddened by E(B-V) $\sim 0.12$, and the
H$\alpha$/H$\beta$ ratio even more reddened with E(B $-$ V) $\sim$\
0.45.  The broad emission lines are unpolarized.  We see broad Na I
$\lambda$5892 absorption in the unpolarized continuum, but not in
the polarized continuum (at the 5-6 $\sigma$\ level).  The polarized
continuum rises smoothly towards shorter wavelengths with $F_{\lambda}
\propto \lambda^{-2}$.  We argue that a
normal QSO continuum is polarized by scattering from a region within,
or very near, the Broad Emission Line Region (BELR).  Thus there are
at least three distinct light paths to the observer: a dusty path from
the BELR, a direct path traced by the unpolarized continuum, passing
through dust and low-ionization gas (NaI), and another relatively
unobscured path followed by scattered continuum.  This provides direct
evidence that a BAL region and dust only partially cover the central QSO.

Ultraluminous AGNs, including IRAS 07598+6508, appear no more
IR-luminous than non-IRAS-selected QSOs,
and have normal L$_{\rm IR}$/L$_{\rm opt}$ ratios when the optical
luminosities are corrected for reddening.
Reddening and BALs occur only along some sight lines and the parent
population of BALQs are `normal' QSOs.

\end{abstract}
\keywords{quasars: absorption lines --- quasars: emission lines ---
quasars: general --- quasars: individual (IRAS 07598+6508) --- polarization}
\vfil
{\centerline {Submitted to The Astrophysical Journal, Part 2}}
\vfil\eject
\section{Introduction}

There are some very clear observational relationships among various
properties of luminous AGNs that are important to investigate, because
they bear on the inter-related themes of unification by orientation,
by covering factors and amounts of cool and low-ionization material,
and by the power of radio jets.  The strongest relationships are the
inverse correlation between optical Fe II and [OIII] $\lambda$5007
strength (Boroson \& Green 1992, Boroson \& Meyers 1992), and the
extremely strong Fe II (optical) emission and strong broad absorption
lines (BALs) found only among radio-quiet QSOs (Stocke et al. 1992,
Lipari, Terlevich, \& Macchetto 1993).

Among radio-loud quasars, orientation unification supposes that
lobe-dominant quasars represent objects whose central engine axis
is highly inclined.
Core-dominated blazars represent the low-inclination
extreme.  In the FR II radio galaxies, the jet axes are closest to the
sky plane, and the QSO is buried within a dusty torus whose axis is
aligned with the central engine's.  In lobe-dominated sources,
non-synchrotron, optical polarization is often found aligned either
parallel or perpendicular to the jets.  In some well studied cases
this is attributed to central QSO emission scattered into the
line-of-sight, and in some cases the scattering region is resolved --
in continuum images it is seen aligned with the radio jets (di Serego
Alighieri, Cimatti, \& Fosbury 1993).  The more lobe-dominated sources
have weaker [OIII] $\lambda$5007 emission -- attributed to shadowing
of the higher-ionization NLR by a dusty torus aligned with the jet
(Hes, Barthel, \& Fosbury 1993).  At least in some cases the continua
and emission lines of lobe-dominant sources appear reddened.  Also,
associated absorption may be more common in lobe-dominated quasars
(Wills et al. 1995; Aldcroft, Elvis, \& Bechtold 1995).

Among radio-quiet QSOs, the following appear to be related:
low-ionization BALs, super-strong Fe II emission, extremely weak
[OIII] $\lambda$5007 emission, reddened spectral energy distributions,
and non-blazar linear polarization.  Despite the above relations, the
ultraviolet emission line spectra of non-BAL and BAL QSOs are,
overall, very similar, and so it was suggested that BAL differences
might arise simply as a result of different viewing angles (Weymann et
al. 1991; Hartig \& Baldwin 1986); a disk geometry for the BALR was
proposed (Turnshek 1988).  Could some of the above similarities between
lobe-dominated quasars and BAL QSOs be attributed to the same
axisymmetric model for the inner few parsecs?

The discovery by Wills et al. (1992) of high, wavelength-dependent
polarization in the first IRAS-discovered QSO, IRAS 13349+2438, led
them directly to the now-standard explanation -- the observed spectrum
is the combination of QSO light reddened by passage through a dusty
torus, and less-reddened polarized light scattered from within the
opening of the torus.  At the time, BAL QSOs were the only QSOs
showing significant non-blazar polarization (Stockman, Moore \& Angel
1984).  Wills et al. (1992) therefore predicted that significant
polarization and BALs might be found in other IRAS-selected QSOs,
their model for IRAS 13348+2438 naturally leading to the idea that
normal QSOs could appear as non-BAL or BAL QSOs depending on the
viewing angle.  Our subsequent polarization survey of the Low et
al. (1988) sample of IRAS-selected QSOs, to investigate unification
schemes and the r\^ole of dust, led to the discovery of significant
polarization in most of the sample, including IRAS 07598+6508 (Wills
\& Hines 1995, see also Hines \& Wills 1993).  Our IUE spectroscopy of
IRAS 07598+6508 (Fig. 1) revealed the predicted BALs.

In this {\it Letter} we present spectroscopy and spectropolarimetry
of IRAS 07598+6508 and use these results
to constrain an anisotropic geometry for the scattering and absorbing regions.
IRAS 07598+6508 is the only QSO we know of that embodies all the above
radio-quiet characteristics in an extreme way, and our new results show
how this QSO may provide an important link between apparently different
observational classes of QSOs.

\section{Results}

We used the International Ultraviolet
Explorer (IUE) to obtain two SWP and two LWP spectra, with particular
care in centering in the 20 $^{\prime\prime}$\ aperture, to obtain
reliable wavelength calibration and spectrophotometry.  Reduction was
by NEWSIPS
with optimal extraction techniques, using the duplicate
observations to remove cosmic ray `hits' and improve the s/n.  The
same data have been discussed by Lipari (1994).  The ground-based data
were obtained and reduced as described by Hines \& Wills (1993).

Figure 1 presents the UV-optical spectrophotometry of IRAS 07598+6508.
Note the rich Fe II spectrum originally found by Lawrence et al. (1988),
and the absence of significant NLR emission.  (The optical spectrum
has been discussed by Boroson \& Meyers 1992, who were the first to
recognize the broad Na I $\lambda$5892 absorption trough in their
higher resolution spectrum; Lipari 1994).  The low-ionization BALs are
among the strongest known, with EW (Mg II $\lambda$2798) = 62 \AA, but
show trends seen in other low-ionization BAL QSOs (Voit et al. 1993).
Using the peak of the broad H$\,\alpha$ emission line to define the
rest frame, the BAL troughs extend between blue shifts of 5,200 and
22,000 km s$^{-1}$, as shown by the horizontal lines on the
figure.  As in other BAL QSOs, the peaks of the high-ionization UV
lines are blueshifted with respect to H $\alpha$, but in IRAS
07598+6508 this shift is especially large -- 3000 km s$^{-1}$\
relative to the Balmer line and Na I $\lambda$5892 emission peaks.
The H$\,\alpha$/H$\,\beta$ intensity ratio of 6.2 $\pm $\ 0.8, when
compared with typical values for UV-selected QSOs (3.3, Thompson
1992), indicates broad emission line reddening of E(B $-$ V) $\sim 0.45$.
As in other low-ionization BAL QSOs, the Balmer line EWs are low, with
EW(H $\alpha$) $\sim 275$\ \AA.  The spectral energy distribution is
less reddened, and matches that of a typical QSO, if an SMC reddening
curve is adopted with E(B $-$ V) $\sim$\ 0.12; this value is typical
for the UV spectra of other low-ionization BAL QSOs
(Sprayberry \& Foltz 1992).

The spectropolarimetry results are shown in Figure 2.  The percentage
polarization (represented by a Stokes parameter rotated to the
wavelength-independent position angle of 116$^{o}$\ shown in the
bottom panel) decreases in regions of line emission and increases to
about 3.5\% in the region of the Na I $\lambda$5892 BAL.  The third
panel shows the corresponding polarized flux-density spectrum. This is well
fitted by $F_\lambda \propto
\lambda^{-2}$, showing no significant features, even in the region of the Na I
BAL.  Blended Fe II emission contributes increasingly at the shorter
wavelengths, resulting in the decline of percentage polarization
shortward of 4500 \AA.

The polarization in the broad emission lines was derived by
subtracting a smoothed continuum from each Stokes flux-density
spectrum (Q$_{\lambda}$ and U$_{\lambda}$), using the regions of
minimum line contribution, to yield Q$_{\lambda}$\ and U$_{\lambda}$\
for the emission lines alone.  By integrating the Q$_{\lambda}$\ and
U$_{\lambda}$\ over the emission lines and dividing by the observed
total line intensities, we derive an emission-line polarization of 0.16 $\pm
0.04$\% and position angle $\theta = 150^{o}$.  This is less
than the maximum expected $0.5$\% Galactic interstellar polarization.

In order to investigate the continuum polarization, we have
subtracted the Fe II blends and other broad emission lines, using the
method and I Zw 1 template of Boroson \& Green (1992) (for
$\lambda > 4300$\AA, upper panel of Fig. 3).
The lower panel of Fig. 3 shows the polarization
spectrum after subtracting the unpolarized emission lines.  Between
6800 \AA\ and 4300 \AA\ the polarization rises from 2 to 3\%, except
in the region of the Na I BAL where it rises to 3.5 $\pm 0.3$\%.
This peak is entirely accounted for by absorption in the unpolarized
continuum.  Such absorption is absent from the polarized flux
density spectrum at a significance level of 5 -- 6 $\sigma$.  The
maximum intrinsic polarization of the polarized continuum must be at least 3.5
$\pm 0.3$\%.

\section{The Geometry and Polarization Mechanism}

The spectropolarimetry shows that there are at least three spectral components:
(i) light from the
broad emission line region (BELR) that is unpolarized and reddened with E(B
$-$ V) $\sim $\ 0.45, (ii) unpolarized continuum that passes through a
low ionization BALR, and also suffers significant reddening, E(B $-$
V) $> 0.12$,  (iii) polarized continuum that is much less reddened
than component (ii) and does not pass through significant Na I BAL material.
We adapt a standard QSO geometry in which the BELR
lies at $\sim$ 0.1 to 1 pc from, and partially covers, a central
continuum source.

The simplest explanation for the polarized continuum would be
synchrotron emission, e.g., from the inner regions of a stable jet
(our broad-band polarization observations over three years show no
signs of variability).  However, the
observed rise of the polarized flux density spectrum toward shorter
wavelengths has never been seen in spectra attributed to
synchrotron radiation.  Also, it would be barely possible to explain
the strength of the optical polarized flux density given IRAS
07598+6508's weak 11.7 mJy radio source (Neff \& Hutchings 1992) and
the flattest observed radio-optical synchrotron spectrum
(as in X-ray-selected BL Lac samples).

If dichroic transmission were the explanation, aligned grains would have to
be within BELR distances of the nucleus or between BELR clouds so that line
radiation would not pass through aligned grains.  The grain properties would be
totally unlike those in our galaxy, with polarization increasing into the
UV (p $> 4$\%).

Thus, we prefer a scattering explanation, and a possible geometry for
the two continuum light paths is shown in Fig. 4.  Note that scattering close
to an accretion disk is excluded because we must account for
unscattered continuum that passes through the BALR.
Any scattered line emission would tend to be unpolarized because of
the almost-symmetric BELR scattering geometry and dilution by direct
line emission.
Figure 1 shows a plausible decomposition of the observed continuum
into a scattered component and direct reddened
component, corresponding to paths a and b in Fig. 4.
With a wavelength-independent polarization of 5\% for the scattered
component and
a reddening for light path `b' equal to the Balmer line reddening, we
can account for the observed wavelength dependence of polarization and
the total spectral energy distribution.

\section{Discussion}

Regardless of the details of the `model' (Fig. 4), different light
paths exist, and some continuum escapes without passing through the
dust and Na I BALR.  Thus IRAS 07598+6508 may be a non-BAL QSO when
viewed from some directions.  Observational characteristics such as
polarization, reddening, extinction, and obscuration are expected to
be aspect-dependent as well.  In FR II narrow-line radio galaxies,
where it has been suggested that an obscuring torus is observed close
to edge-on, Hes et al. (1993) suggest that [OIII] $\lambda$5007 may be
partially hidden.  If this is the entire explanation for the weakness
of [OIII] in BAL QSOs, then the strong Fe II emission is dependent
on viewing angle, even more so than other broad emission lines.

It might be expected that QSOs selected by IRAS flux density would be
biased towards those with especially high L$_{IR}$, so that the
existence of reddening and BALs among the IRAS QSOs could be
attributed to larger covering by dust and low-ionization gas rather
than to orientation.  Such does not seem to be the case. If we correct
the ratio of infrared to optical luminosity for reddening (Hines \&
Wills 1995), we find that the IRAS-discovered QSOs, including IRAS
07598+6508, are indistinguishable from the PG QSOs in an L$_{\rm
IR}$/L$_{\rm opt}$ vs. L$_{\rm bol}$ diagram (Fig. 2 in Low et
al. 1989; Fig. 4 in Cutri et al. 1994).  The IRAS-discovered QSOs were
not known previously, almost certainly because their UV-optical spectra
are too red to be selected by UV excess (Hines \& Wills 1995, Wills et
al. 1992).  The similarity in L$_{IR}$, combined with the fact that
the de-reddened continuum and Balmer line ratios and the scattered
spectra resemble those of typical PG QSOs, strongly supports
the suggestion that all QSOs harbor a BALR, but some are viewed
from a direction that intercepts the BAL clouds.

We have no information concerning any fundamental axis in IRAS
07598+6508, although faint optical extensions have been observed,
suggesting tidal interaction (Sanders, private communication).
The strongest support that we can muster for
an axisymmetric model is by analogy with the
ionization and scattering cones and jet directions observed for some
lower luminosity polarized AGNs.  IRAS 07598+6508 has been selected by
its 60 $\mu$m flux density and ``warm'' IRAS colors in a way similar
to other polarized IRAS QSOs, and at least in the case of IRAS
13348+2408, where a similar scattering model has been proposed, the
polarization position angle is aligned with the host galaxy's major
axis.

An important difference between IRAS 07598+6508 and other polarized
AGNs is the lack of significant polarization of the emission lines.
Glenn, Schmidt, \& Foltz (1994) find the same for the BAL QSO CSO755
(see also PHL 5200 -- Stockman, Angel, \& Hier 1981; Goodrich et
al. 1995).  (Note that neither of these objects was selected by a
bright, UV excess continuum.)  Perhaps this difference could be
attributed to orientation by means of angular dependence of optical
depth and the grain scattering function.  Antonucci (1988) reports
this phenomenon in three low-polarization non-blazar radio-loud
quasars, suggesting that the `model' for IRAS 07598+6508 may be
applicable to radio-loud quasars.

If the polarized continuum in IRAS 07598+6508 does arise from scattering, the
scattering region is within or near the BELR, suggesting that photoionization
models for the BELR should take into account these scattered continuum
photons either within the BELR, or by external illumination (Kallman \&
Krolik 1986).

\acknowledgments

Thanks to the staff of McDonald Observatory for willing help,
especially to David Doss and Ed Dutchover.  We thank Bob Goodrich for
fitting the polarization optics into the Large Cass Spectrograph, with
help from David Boyd.  Grant support for this was via University of
Texas University Research Institute grant URI R-154 and a
Summer Research Award (B. J. W.).  Bob Goodrich provided valuable
advice on VISTA data reduction.  It is a pleasure to thank Jean
Clavel, Ron Pitts, Cathy Imhoff, Randy Thompson, and the TAs at Goddard
and VILSPA (Spain); also Willem Wamsteker and Yoji Kondo who made the
IUE observations possible by special scheduling.  We thank T. Boroson
for the IZw1 FeII template, and Derek Wills for making some early
observations.  B. J. W. is grateful to NASA for IUE grant NAG 5-1108,
and B. J. W. and D. C. H. to the Space Telescope Science Institute
for grant GO-5463.

\eject
\centerline{\bf Figure Captions}

\noindent
FIG. 1.\ The combined IUE-McDonald Observatory total flux density
spectrum.  $F_{\lambda}$\ is in units of $10^{-16}$ erg s$^{-1}$
cm$^{-2}$ \AA$^{-1}$.  Emission lines are identified by vertical lines
at the wavelengths predicted from the MgII redshift, and corresponding
absorption troughs are indicated by horizontal lines.  The dashed line
represents our best estimate of the underlying continuum, a power law
with reddening like the Balmer lines. The dotted line represents our
model of the scattered continuum (see text).

\noindent
FIG. 2.\ Spectropolarimetry results.  (a) the total flux density spectrum
($F_{\lambda}$\ in units of $10^{-16}$ erg s$^{-1}$ cm$^{-2}$ \AA$^{-1}$),
corrected for a Galactic extinction of E(B $-$ V) = 0.06, (b) percentage
polarization (represented by RSP$_\lambda$), (c) polarized flux density
(represented by SF$_\lambda$) in same units as (a), and (d)
polarization position angle $\theta_\lambda$ (degrees).

\noindent
FIG. 3.\ (a) FeII deblending.  The total flux density spectrum (dotted
line), the FeII-subtracted spectrum (solid line), power-law continuum
fit ($\alpha_\nu \approx -0.4$: dashed), and the FeII template (bottom
solid line).  F$_\lambda$ is in the same units as in Fig. 1.  (b) the
continuum polarization corrected for dilution by the strong,
essentially unpolarized, line emission.  Note the strong increase in
polarization across the NaI BAL showing that the NaI absorption is
present only in unpolarized light.

FIG. 4.\ A representation of the scattering geometry. (a) is the
path inferred for the scattered light, and (b) is the direct, reddened
path, which also passes through broad absorbing material.

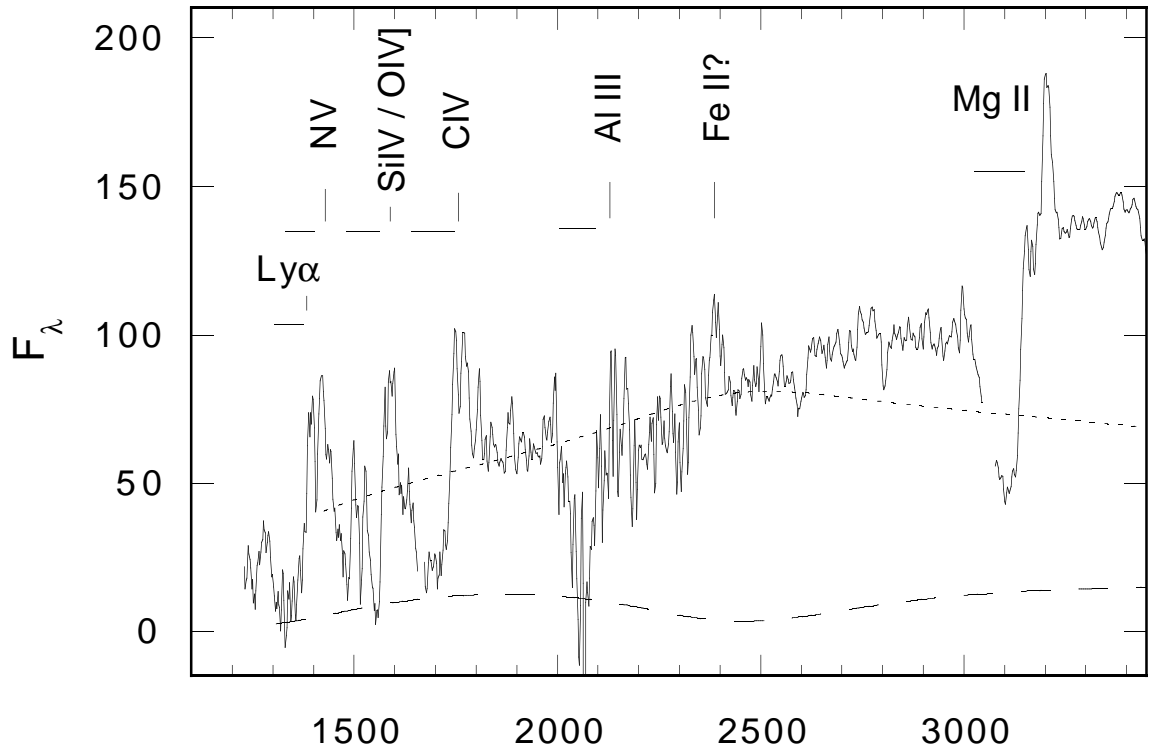
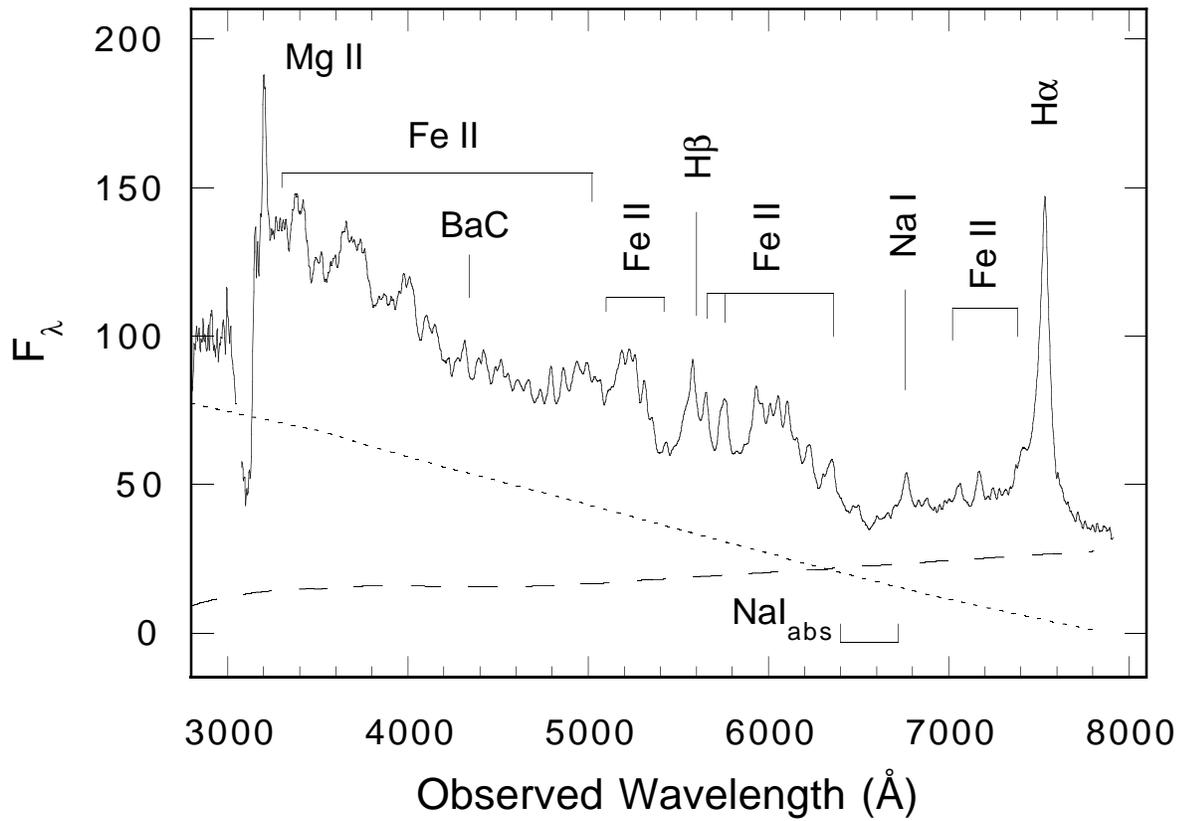
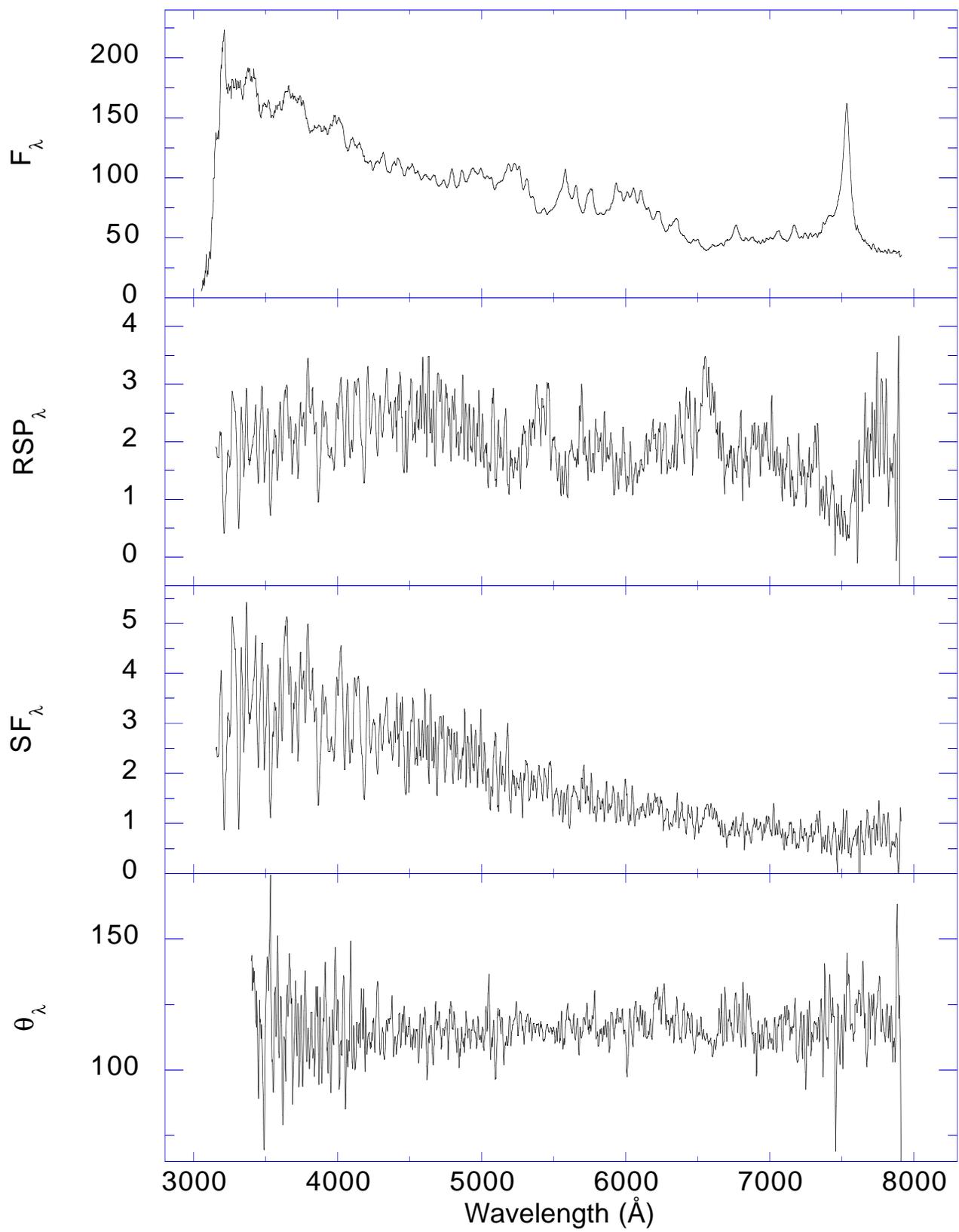
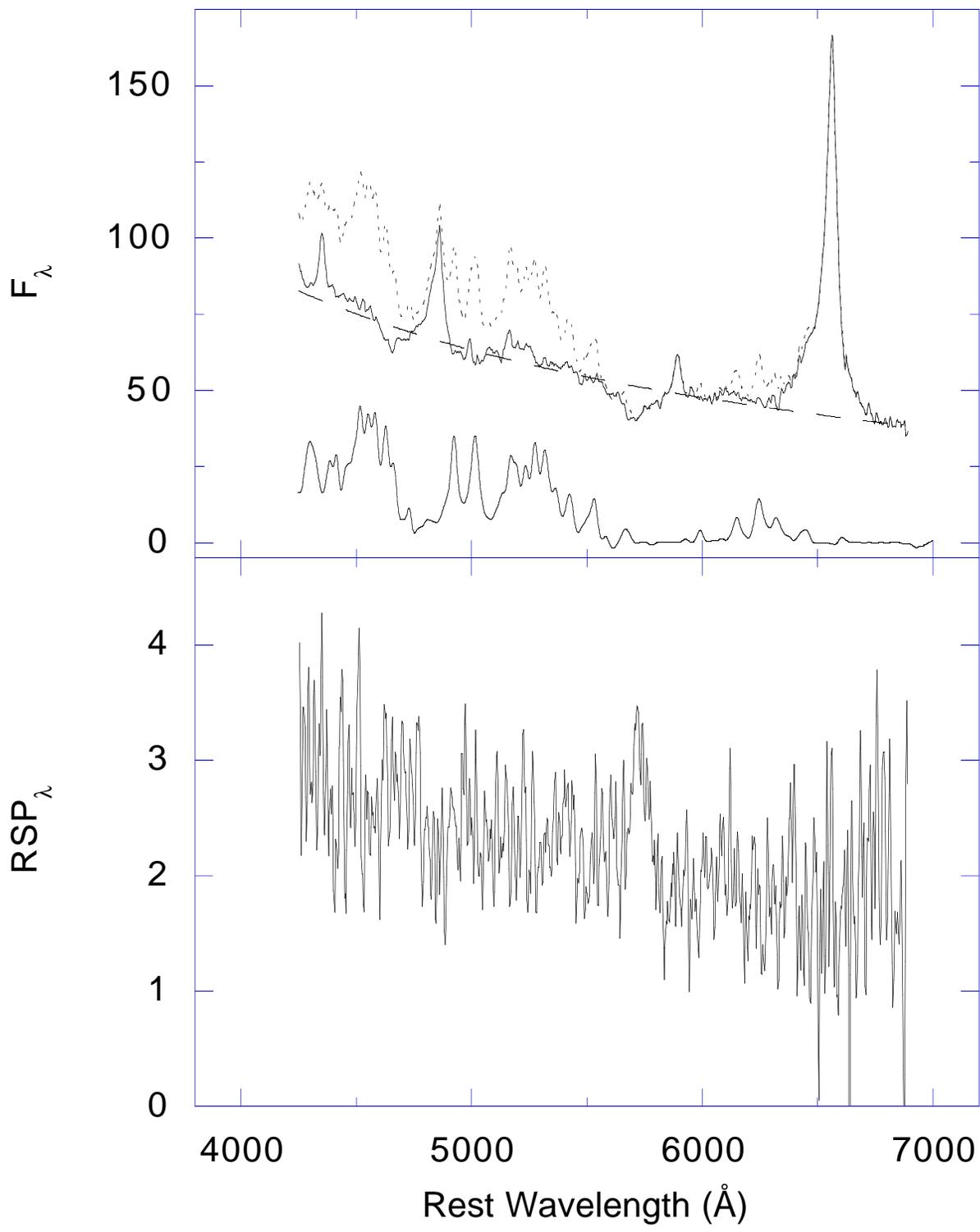
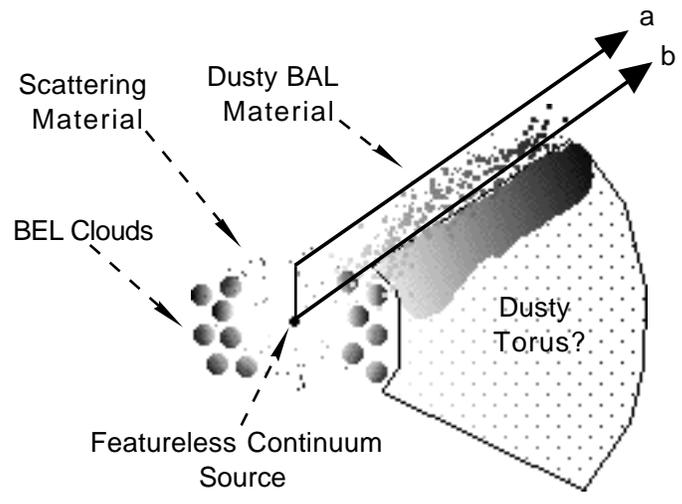
\end{document}